\documentclass[pra, aps,showpacs,amsmath]{revtex4}
\def\mZ{\mathbb{Z}}
\def\mF{\mathbb{F}}

\def\cH{\mathcal{H}}
\def\cK{\mathcal{K}}
\def\hW{\widehat{W}}
\def\cN{\mathcal{N}}
\usepackage{amssymb}
\usepackage{amsfonts}
\usepackage{amsmath}
\begin{document}
\title{Wigner distributions for finite state systems without redundant phase point operators}
\author{S. Chaturvedi}
\email{scsp@uohyd.ernet.in}
\affiliation{School of Physics, University of Hyderabad,
Hyderabad 500046}
\author{N. Mukunda}
\email{nmukunda@cts.iisc.ernet.in}
\affiliation{Centre for High Energy Physics, Indian Institute of
Science, Bangalore 560012}
\author{R. Simon}
\email{simon@imsc.res.in}
\affiliation{The Institute of
 Mathematical Sciences, C. I. T. Campus, Chennai 600113}

\begin{abstract}
We set up Wigner distributions for $N$ state quantum systems
following a Dirac inspired approach. In contrast to much of the
work on this case, requiring  a $2N\times 2N$ phase space, particularly 
when $N$ is even, our approach is
uniformly based on an $N\times N $ phase space grid and thereby
avoids the necessity of having to invoke a `quadrupled' phase
space and hence the attendant redundance. Both $N$ odd and even
cases are analysed in detail and it is found that there are
striking differences between the two. While the $N$ odd case
permits full implementation of the marginals property, the even
case does so only in a restricted sense. This has the consequence
that  in the even case one is led to several equally good
definitions of the Wigner distributions as opposed to the odd case
where the choice turns out to be unique.
\end{abstract}
\pacs{03.65.-w; 03.65.Ca; 03.65.Wj\\
{\bf Keywords:} Wigner distributions, Phase Space, Isotropic lines, Phase Point Operators, Marginals Property, Symplectic Group  $SL(2,\mZ_{N})$, Tomographic Reconstruction.}

\maketitle

\section{Introduction}
The Wigner distribution, as originally defined by Wigner \cite{1} for the case of a Cartesian configuration space has, over the years,
been extended and generalised in many directions \cite{2}-\cite{a}. These extensions and generalisations (most of which have been elegantly unified in a recent work of Ferrie and Emerson \cite{F} using the language of frame theory) include situations where the coordinates take values in
\begin{itemize}
\item a finite field $\mF_{p^n}$ \item a ring  $\mZ_N$ \item a
finite group (abelian or non abelian) \item the manifold of a semi
simple compact Lie group
\end{itemize}
The first two are of particular interest owing to potential applications to quantum information processing and quantum state estimation.
In the present work we focus our attention on the second case and set up the phase space and Wigner distributions  thereon using what we call
 a Dirac inspired approach to Wigner distributions \cite{13}. This approach essentially requires computing the square root of a certain kernel
which brings with it undetermined signs, one at each phase point.
We examine the question to what extent these signs can be fixed or
related to each other by the marginals property i.e by demanding
that the phase space averages of the Wigner distribution along an
`isotropic line' and the lines `parallel' to it  yield a
probability distribution. For the case when $N$ is odd one finds
that one can consistently impose marginals property on all
`isotropic lines' which in turn uniquely fixes all the signs. The
Wigner distribution thus obtained turns out to be the same  as
that already known in the literature. For the even case, on the
other hand, this is not so. We find that the marginals property
can not be consistently imposed on all the `isotropic lines' but
only on specific subsets thereof (i.e. on individual orbits under
$SL(2,\mZ_N) )$. Therefore the best one can do is to  demand the
marginals property on the largest such subset. This is actually
good enough as the `isotropic lines' in this subset ( and only in
this subset) cover all phase points. We explicitly carry  this out
and find that, unlike the odd case, not all signs get fixed and
therefore one has a family of Wigner distributions characterised
by different choices for the signs , all of which respect the
restricted marginals property. We emphasise that here we work all
along with an $N\times N$ phase space lattice and the associated $N^2$ phase point operators. This is  in contrast to
other formalisms to be found in the literature \cite{2}, \cite{3} where a $2N\times
2N$ phase space lattice is invoked to arrive at  a satisfactory definition
of Wigner distributions for an $N$-state quantum system. An interesting recent
work by Bar-on \cite{a} based on symmetric informationally complete projection operator valued measures (SIC-POVMs) \cite{b} and a block design theory \cite{c} inspired picture of phase space, reduces the number 
phase points from $4N^2$ to $N^2+N$, but still has $N$ phase points more than those required in the present work. 

A brief outline of this work is as follows: In Section II we set
up our notation and summarise the properties of the operators used
later. In Section III we briefly recapitulate the Dirac inspired
approach to Wigner distributions and give the expressions for the
phase points operators in a convenient form along with the
conditions on the undetermined signs that appear therein in order
that the phase point operators satisfy the standard marginals
property. In Section IV we briefly summarise the properties of the
isotropic lines in our phase space as made available in a recent
work by Albouy\cite{14}. In Section V we discuss the odd case and
in Section VI summarise our result for the case when $N$ is a prime power. The case of general $N$ is discussed inSection VII. In Section VIII we discuss the question of tomographic reconstruction for the case when $N=2^n$. We conclude with Section IX 
 where we note that although for a
given even $N$ there is a great variety of Wigner distributions
characterised by different choices for the unfixed signs, the
actual number of choices available, in so far as the eigenvalues
of the relevant phase point operators is concerned, is rather
small.

\section{Preliminaries}
Consider a quantum system described by a complex Hilbert space
${\cal H}$ of dimension $N$. Denote by $\{|q\rangle\},\{|p)\},
q,p\in \mZ_N$ two orthonormal bases - coordinate and momentum
bases - related to each other by a finite Fourier transform :
\begin{eqnarray}
\langle q^\prime|q\rangle=\delta_{q^\prime,q};~(p^\prime|p)=\delta_{p^\prime,p};~\langle q|p)=\omega^{qp}/\sqrt{N};~~\omega=e^{2\pi i/N}
\end{eqnarray}
Here $\mZ_N$ stands for the ring of integers $\{0,1,2,\cdots,N-1\}$ with addition and multiplication modulo $N$. We use the notation $[n]$ to
denote $n$ modulo $N$. On the Hilbert space $\cal{H}$ we introduce the familiar
Weyl operators:
\begin{eqnarray}
&&U=\omega^{\hat{q}}=\sum_{q\in Z_N}\omega^q|q\rangle\langle q|,~U|p)=|[p+1]);~~U^\dagger U=U^N=1.\nonumber\\
&&U^p=\omega^{p\hat{q}}=\sum_{q\in Z_N}\omega^{pq}|q\rangle\langle q|;~|q\rangle\langle q|=\frac{1}{N}\sum_{p\in Z_N}\omega^{-qp}U^p.\\
&&V=\omega^{-\hat{p}}=\sum_{p\in Z_N}\omega^{-p}|p)( p|,~V|q\rangle=|[q+1]\rangle;~~V^\dagger V=V^N=1\nonumber\\
&&V^q=\omega^{-q\hat{p}}=\sum_{p\in Z_N}\omega^{-pq}|p)(p|;~|p)(p|=\frac{1}{N}\sum_{q\in Z_N}\omega^{qp}V^q.
\nonumber\\
&&U^pV^q=\omega^{pq}V^qU^p=\tau^{2pq}V^qU^p,
\end{eqnarray}
where $\tau=e^{\pi i/N}$.
We denote by $\Gamma_0$ the discrete `classical' phase space of $N^2$ points $\sigma=(q,p)$ equipped with the symplectic product
 $ \langle\sigma,\sigma^\prime\rangle=pq^\prime-qp^\prime$ between any two phase points $\sigma$ and $\sigma^\prime$. We denote by ${\cal K}$
 the Hilbert space of square summable functions on $\Gamma_0$:
\begin{eqnarray}
\cK&=&L^2(\Gamma_0)=N^2{\rm -dimensional~Hilbert~space}\nonumber\\
&=&\{f(\sigma)\in C~|~\sigma\in\Gamma_0,~\parallel f\parallel^2=\sum_{\sigma\in\Gamma_0}~|f(\sigma)|^2\}.
\end{eqnarray}
$\Gamma_0$ can be viewed as an abelian group of order $N^2$:
\begin{eqnarray}
&&\sigma\in\Gamma_0\rightarrow {\rm group~element}~ g(\sigma)=a^qb^p,\nonumber\\
&&{\rm generators}~a,b:~ab=ba,~a^N=b^N=e\nonumber\\
&&g((0,0))={\rm identity}~e;\label{5}\\
&&{\rm composition}:g(\sigma^{\prime})g({\sigma})=g(([q^\prime+q],[p^\prime+p]))=g([\sigma+\sigma^\prime]);\nonumber\\
&&{\rm inversion}:~g(\sigma)^{-1}=g(([N-q],[N-p]))\equiv g([N-\sigma]).\label{6}
\end{eqnarray}
Thus $\Gamma_0$ is the direct product $G_0\times G_0$ where $G_0$
is the (abelian) cyclic group of order $N$.

 $\Gamma_0$ has $N^2$
distinct inequivalent one--dimensional unitary irreducible
representations (UIR's), with characters labelled by points
$\sigma^\prime\in \Gamma_0$ :
\begin{eqnarray}
&&\sigma^\prime\in \Gamma_0:~{\rm UIR}~~a\rightarrow \omega^{-p^\prime},~b\rightarrow \omega^{q^\prime};
\nonumber\\
&&g(\sigma)\rightarrow\chi_{\sigma^\prime}(\sigma)=\omega^{pq^\prime-qp^\prime}=\omega^{\langle \sigma,\sigma^\prime\rangle};\nonumber\\
&&{\rm Orthogonality}:\chi_{\sigma^{\prime\prime}}^\dagger\chi_{\sigma^{\prime}}=\sum_{\sigma\in\Gamma_0} \chi_{\sigma^{\prime\prime}}^*(\sigma)\chi_{\sigma^{\prime}}(\sigma)\nonumber\\
&&~~~~~~~~~~=\sum_{q,p\in \mZ_N}\omega^{p^{\prime\prime}q-q^{\prime\prime}p+q^{\prime}p-p^{\prime}q}=
N^2\delta_{\sigma^{\prime\prime},\sigma^{\prime}}
\end{eqnarray}
The set of functions $\{\frac{1}{N}\chi_{\sigma^{\prime}}(\sigma)\}$ on $\Gamma_0$ forms an orthonormal basis (ONB) for ${\cal K}$. The trivial UIR is $\sigma^{\prime}=0, \chi_0(\sigma)=1$, and hence
\begin{equation}
 \sum_{\sigma\in\Gamma_0}\chi_{\sigma^{\prime\prime}}(\sigma)=\sum_{\sigma\in\Gamma_0}\omega^{\langle\sigma,
\sigma^{\prime\prime}\rangle}=N^2\delta_{\sigma^{\prime\prime},0}
\end{equation}
Finally we introduce the displacement operators $D(\sigma)$ on $\cal{H}$, one for each phase point, and summarise their properties:
\begin{eqnarray}
 &&\sigma\in\Gamma_0:~D(\sigma)\equiv D(q,p)=\tau^{qp}V^qU^p=\tau^{-qp}U^pV^q;\\
&&{\rm Unitarity}:~D(\sigma)^\dagger D(\sigma)=1~{\rm on}~\cH.\\
&&{\rm Inverses}:~D(\sigma)^{-1}=\eta_{\sigma}D([N-\sigma]),\\
&&~~~~~~~~~~~~~~~~~~~~~\eta_{\sigma}=\begin{cases}1~{\rm for}~q=0~{\rm or}~p=0 \\
(-1)^{q+p+N}~{\rm for}~1\leq q,p\leq N-1\end{cases}\nonumber
\\&&~~~~~~~~~~~~~~~~~~~~~~~~=\tau^{-[N-q][N-p]+qp}\\
&&{\rm Composition}:~D(\sigma^\prime)D(\sigma)=\tau^{\langle\sigma^\prime,\sigma\rangle}\epsilon(\sigma^\prime,\sigma)
D([\sigma^\prime+\sigma]),\\
&&~~~~~~~~~~~~~~~~~~~~~~\epsilon(\sigma^\prime,\sigma)
=\epsilon(\sigma,\sigma^\prime)=\begin{cases}~1~{\rm for}~\sigma^\prime+\sigma\in\Gamma_0\\
(-1)^{q^\prime +q}~{\rm for}~q^\prime+q\leq N-1,p^\prime+p\geq N\\
(-1)^{p^\prime +p}~{\rm for}~q^\prime+q\geq N,p^\prime+p\leq N-1\\
(-1)^{q^\prime +q+p^\prime+p+N}~{\rm for}~q^\prime+q\geq N,p^\prime+p\geq N\end{cases}\nonumber\\
&&~~~~~~~~~~~~~~~~~~~~~~~~~~~~~~~~~~~~~~~~~~~~=\tau^{(q^\prime+q)(p^\prime+p)-[q^\prime+q][p^\prime+p]}
\label{18}\\
&&{\rm Trace~orthogonality}:
~~~{\rm Tr}(D(\sigma^\prime)^\dagger D(\sigma))=N\delta_{\sigma^\prime,\sigma}
\end{eqnarray}

The set of operators $\{\frac{1}{\sqrt{N}}D(\sigma)\}$ constitutes a complete irreducible trace orthonormal set of operators on $\cH$ satisfying the relations:
\begin{eqnarray}
&&D(\sigma^\prime) D(\sigma)=\omega^{\langle \sigma^\prime,\sigma\rangle} D(\sigma)D(\sigma^\prime)\label{21}\\
&&D(\sigma^\prime) D(\sigma)D(\sigma^\prime)^{-1}=\omega^{\langle \sigma^\prime,\sigma\rangle} D(\sigma)
\label{22}
\end{eqnarray}
Thus $\{D(\sigma)\}$ form an irreducible unitary $N$ dimensional ray representation of $\Gamma_0$ on $\cH$.  Some useful relations are given below:
\begin{eqnarray}
&&\sigma\in \Gamma_0:~\tau^{-qp}D(\sigma)=V^qU^p=\sum_{\sigma^\prime\in\Gamma_0}\omega^{\langle \sigma,\sigma^\prime\rangle}|p^\prime)(p^\prime|q^\prime\rangle\langle q^\prime|,\\
&&|p)(p|q\rangle\langle q|=\frac{1}{N^2}\sum_{\sigma^\prime\in\Gamma_0}\omega^{\langle \sigma,\sigma^\prime\rangle}\tau^{-q^\prime p^\prime}D(\sigma^\prime).
\label{24}
\end{eqnarray}
\section{Dirac inspired approach to Wigner distributions}
The central idea in this approach, discussed in detail in
\cite{13}, is to initially associate with any operator
$\widehat{A}$ a phase space function $A(\sigma)$ constructed out
of its mixed matrix elements such as $\langle q|\widehat{A}|p)$ in
such a way that the trace of the product of two operators is
expressed as a phase space average of a kernel times the product
of their phase space functions, and then try to transform away
this kernel. As shown in \cite{13}, this exercise entails finding
a symmetric square root of the kernel $K(\sigma,\sigma^\prime)$
defined below:
\begin{equation}
K(\sigma;\sigma^\prime)=K(\sigma^\prime;\sigma)=\omega^{(q-q^\prime)(p-p^\prime)}.
\end{equation}
\subsection{ The~ Kernel~ K and its symmetric square roots $\xi$}
The  kernel $K$ defines an operator on $\cK$. The characters $\{\chi_{\sigma^\prime}(\sigma)\}$ of $\Gamma_0$ form a complete orthogonal set of eigenvectors of $K$ in $\cK$:
\begin{equation}
\sum_{\sigma^\prime \in \Gamma_0} K(\sigma;\sigma^\prime)\chi_{\sigma^{\prime\prime}}(\sigma^\prime)=
N\omega^{q^{\prime\prime}p^{\prime\prime}}\chi_{\sigma^{\prime\prime}}(\sigma)
\end{equation}
Using the results above for the eigenvalues and eigenvectors of $K$, a general square root $\xi$ of $K$ is defined by
\begin{equation}
\xi\chi_{\sigma^{\prime\prime}}=\sqrt{N}\tau^{q^{\prime\prime}p^{\prime\prime}}S(\sigma^{\prime\prime})\chi_{\sigma^{\prime\prime}},~S(\sigma^{\prime\prime})=\pm 1,~\sigma^{\prime\prime}\in \Gamma_0.
\end{equation}
At this point, there are $N^2$ sign choices. This freedom will be reduced as we proceed. The kernel
$\xi(\sigma;\sigma^\prime)$ is
\begin{eqnarray}
\xi(\sigma;\sigma^\prime)&=&\frac{1}{N^2}\sqrt{N}\sum_{\sigma^{\prime\prime} \in \Gamma_0}\tau^{q^{\prime\prime}p^{\prime\prime}}S(\sigma^{\prime\prime})\chi_{\sigma^{\prime\prime}}(\sigma)\chi_{\sigma^{\prime\prime}}(\sigma^\prime)^*\nonumber\\
&=&\frac{1}{N^{3/2}}\sum_{\sigma^{\prime\prime} \in \Gamma_0}\tau^{q^{\prime\prime}p^{\prime\prime}}S(\sigma^{\prime\prime})\omega^{\langle \sigma,\sigma^{\prime\prime}\rangle-\langle \sigma^\prime,\sigma^{\prime\prime}\rangle}
\end{eqnarray}
$K(\sigma;\sigma^\prime)$ is symmetric under $\sigma \leftrightarrow \sigma^\prime$. Demanding the same for
$\xi(\sigma;\sigma^\prime)$ places conditions on the signs $S(\sigma^{\prime\prime})$. Since in general
\begin{equation}
\sum_{q\in \mZ_N}f(q)=\sum_{q\in \mZ_N}f([N-q])~~{\rm etc},
\end{equation}
\begin{eqnarray}
\xi(\sigma;\sigma^\prime)=\xi(\sigma^\prime;\sigma)\Longleftrightarrow &&
\sum_{\sigma^{\prime\prime}\in\Gamma_0}\tau^{q^{\prime\prime}p^{\prime\prime}}S(\sigma^{\prime\prime})\omega^{\langle \sigma,\sigma^{\prime\prime}\rangle-\langle \sigma^\prime,\sigma^{\prime\prime}\rangle}
\nonumber\\
&&=\sum_{\sigma^{\prime\prime}\in
\Gamma_0}\tau^{q^{\prime\prime}p^{\prime\prime}}S(\sigma^{\prime\prime})\omega^{\langle \sigma^\prime,\sigma^{\prime\prime}\rangle-\langle \sigma,\sigma^{\prime\prime}\rangle}
\nonumber\\
&&=\sum_{\sigma^{\prime\prime}\in
\Gamma_0}\tau^{[N-q^{\prime\prime}][N-p^{\prime\prime}]}S([N-\sigma^{\prime\prime}])\omega^{\langle \sigma,\sigma^{\prime\prime}\rangle-\langle \sigma^\prime,\sigma^{\prime\prime}\rangle}\nonumber\\
\Longleftrightarrow S(\sigma)&=&\tau^{[N-q][N-p]-qp}S([N-\sigma])=\eta_{\sigma}S([N-\sigma]).
\label{30}
\end{eqnarray}
So the number of independent $S(\sigma)$'s is about halved. We
hereafter assume $\xi(\sigma;\sigma^\prime)$ is symmetric.
\subsection{Phase point operators $\hW(\sigma)$} For any choice of
the square root kernel $\xi(\sigma;\sigma^\prime)$, we define
\begin{eqnarray}
\sigma\in\Gamma_0:~\hW(\sigma)&=&\sqrt{N}\sum_{\sigma^\prime\in\Gamma_0}\xi(\sigma;\sigma^\prime)|p^\prime)(p^\prime|
q^\prime\rangle
\langle q^\prime|\nonumber\\
&=&\frac{1}{N}\sum_{\sigma^\prime\in\Gamma_0}\omega^{\langle \sigma,\sigma^{\prime}\rangle} S(\sigma^\prime)D(\sigma^\prime)
\end{eqnarray}
The condition $(\ref{30})$ on $S(\sigma)$ arising from the symmetry of $\xi$  ensures hermiticity of $\hW(\sigma)$:
\begin{eqnarray}
\hW(\sigma)^\dagger&=&\frac{1}{N}\sum_{\sigma^\prime\in\Gamma_0}\omega^{\langle \sigma^\prime,\sigma\rangle} S(\sigma^\prime)D(\sigma^\prime)^{-1}\nonumber\\
&=&\frac{1}{N}\sum_{\sigma^\prime\in\Gamma_0}\omega^{\langle \sigma^\prime,\sigma\rangle} S([N-\sigma^\prime])D([N-\sigma^\prime])\nonumber\\
&=&\hW(\sigma)
\end{eqnarray}
Trace orthogonality of $D(\sigma)$ leads to
\begin{equation}
{\rm Tr}(\hW(\sigma^\prime)\hW(\sigma))=N\delta_{\sigma^\prime,\sigma}
\end{equation}
Further ${\rm Tr}(D(\sigma))=N\delta_{\sigma,0} $ gives
\begin{equation}
{\rm Tr}(\hW(\sigma))=1,
\end{equation}
So both
$\{\frac{1}{\sqrt{N}}\hW(\sigma)\}$ and
$\{\frac{1}{\sqrt{N}}D(\sigma)\}$ are trace orthonormal complete
sets of operators on $\cH$.

From the conjugation relations $(\ref{22})$ for $D$'s we get:
\begin{eqnarray}
D(\sigma^\prime)\hW(\sigma)D(\sigma^\prime)^{-1}&=&\frac{1}{N}\sum_{\sigma^{\prime\prime}\in \Gamma_0}
\omega^{ \langle \sigma,\sigma^{\prime\prime}\rangle}S(\sigma^{\prime\prime})\omega^{\langle \sigma^\prime,\sigma^{\prime\prime}\rangle}D(\sigma^{\prime\prime})\nonumber\\
&=&\hW([\sigma+\sigma^\prime])
\end{eqnarray}
Recovery of standard marginals fixes some $S(\sigma)$:
\begin{eqnarray}
\frac{1}{N}\sum_{p\in \mZ_N}\hW(q,p)&=&\frac{1}{N^2}\sum_{q^\prime,p^\prime,p\in \mZ_N}\omega^{pq^\prime-qp^\prime}
S(q^\prime,p^\prime)D(q^\prime,p^\prime)\nonumber\\
&=&\frac{1}{N}\sum_{p^\prime\in  \mZ_N}\omega^{-qp^\prime}S(0,p^\prime)U^{p^\prime}\nonumber\\
&=&|q\rangle\langle q|\Longleftrightarrow S(0,p^\prime)=1;\label{31}\\
\frac{1}{N}\sum_{q\in \mZ_N}\hW(q,p)&=&|p)(p|\Longleftrightarrow ~S(q^\prime,0)=1\label{32}
\end{eqnarray}
These are consistent with $(\ref{30})$. So at this stage
\begin{equation}
S(q,0)=S(0,p)=1; ~~S(\sigma)=\eta_{\sigma}S([N-\sigma])
\label{33}
\end{equation}
Some useful formulae are given below:
\begin{eqnarray}
{\rm Tr}(\hW(\sigma)D(\sigma^\prime)^\dagger)&=&\omega ^{\langle \sigma,\sigma^{\prime}\rangle}S(\sigma^{\prime}),\nonumber\\
{\rm Tr}(\hW(\sigma)D(\sigma^\prime))&=&\omega ^{\langle \sigma^{\prime},\sigma\rangle}S(\sigma^{\prime}),\nonumber\\
S(\sigma)D(\sigma)&=&\frac{1}{N}\sum_{\sigma^\prime\in \Gamma_0}\omega^{ \langle \sigma,\sigma^{\prime}\rangle}W(\sigma^\prime)
\end{eqnarray}
\section{Isotropic~lines~and~further~marginals} The conditions so far on
$S(\sigma)$ are given above in $(\ref{33})$. To generate more
conditions, we consider more marginals conditions, based on
isotropic lines.

An isotropic line $\lambda$ is a maximal set of $N$ distinct points in $\Gamma_0$ including
$\sigma=(0,0)$
and obeying:
\begin{equation}
\sigma^\prime, \sigma \in\lambda\Rightarrow \langle\sigma^\prime,\sigma\rangle=0~{\rm mod}~N.
\end{equation}
( The qualification `mod $N$' will frequently be left implicit).
It is a fact that each point $\sigma\in\Gamma_0$ belongs to at
least one isotropic line. We mention here some useful properties
of such lines, and give further details  in Section VI. The
maximality condition allows us to say :
\begin{equation}
 \sigma\in \Gamma_0,~~\langle\sigma,\sigma^\prime\rangle=0 ~{\rm for~all}~ \sigma^\prime\in\lambda\Rightarrow \sigma\in\lambda.
\end{equation}
This in turn leads to
\begin{equation}
 \sigma\in\lambda\Rightarrow [N-\sigma]\in \lambda,
\label{37}
\end{equation}
where $[N-\sigma]$ is defined in $(\ref{6})$. We also have closure under the group composition law
$(\ref{5})$,~$(\ref{6})$ in $\Gamma_0$ :

\begin{equation}
\sigma^\prime, \sigma \in\lambda\Rightarrow [\pm 2\sigma], [\pm 3\sigma],\cdots,[\sigma+\sigma^\prime]\in \lambda.
\label{38}
\end{equation}

In fact the points $\{\sigma\}$ of $\lambda$ form an (abelian) subgroup of $\Gamma_0$, of order $N$, with group composition being (component--wise) addition ${\rm mod}~ N$ as in eqs. $(\ref{5})$,~$(\ref{6})$; the content is the same as in eq. $(\ref{38})$:
\begin{equation}
\sigma^\prime, \sigma \in\lambda\rightarrow g(\sigma^\prime)g(\sigma)=g([\sigma^\prime+\sigma]),~[\sigma^\prime+\sigma]\in \lambda
\end{equation}
From this group structure we see that if any $\sigma, ~[\sigma^\prime+\sigma] \in \lambda$ are given, then $\sigma^\prime \in \lambda$ is uniquely determined. In case $g(\sigma)$ for $\sigma\in\lambda$ is an element of order $N$, $\lambda$ itself is a cycle generated by $\sigma$ and consisting of the $N$ distinct points $\{(0,0),\sigma,[2\sigma],[3\sigma],
\cdots,[(N-1)\sigma]\}$. (However a general $\lambda$ need not be of this form). Examples of such $\sigma$ are
$\sigma=(1,p)$ and $\sigma=(q,1)$.

For any $\sigma^\prime \in \Gamma_0$, we get a (one--dimensional)
unitary irreducible representation (UIR) of $\lambda$  by
\begin{equation}
\sigma\in\lambda \rightarrow \omega^{ \langle\sigma,\sigma^\prime\rangle  }
\end{equation}
If $\sigma^\prime\in \lambda$, this is the trivial UIR . If $\sigma^\prime\notin \lambda$,$\langle\sigma,\sigma^\prime\rangle \neq 0~{\rm mod}~N$ for some $\sigma\in\lambda$, so this is a nontrivial UIR. Hence from orthogonality of inequivalent UIR's we obtain:
\begin{equation}
\sum_{\sigma\in\lambda}\omega^{\langle\sigma,\sigma^\prime\rangle}=N ~{\rm if}~\sigma^\prime\in\lambda,~
0~{\rm if}~\sigma^\prime\notin\lambda
\end{equation}
From the relation $(\ref{21})$ for $D(\sigma)$'s given earlier it
follows that the operators ${D(\sigma),~\sigma\in\lambda}$ form a
mutually commuting set, but they may not form a representation of
$\lambda$. \vskip0.3cm \noindent {\bf
Isotropic~line~marginals~condition} \vskip0.3cm
 Given $\lambda$, define
\begin{eqnarray}
P_\lambda&=& \frac{1}{N}\sum_{\sigma\in\lambda}\hW(\sigma)=\frac{1}{N^2}\sum_{\sigma\in\lambda}\sum_{\sigma^\prime\in\Gamma_0}
\omega^{\langle\sigma,\sigma^\prime\rangle}S(\sigma^\prime)D(\sigma^\prime)\nonumber\\
&=&\frac{1}{N}\sum_{\sigma\in\lambda}S(\sigma)D(\sigma)
\end{eqnarray}
Clearly $P_{\lambda}^\dagger=P_\lambda, {\rm Tr}(P_\lambda)=1$. Now develop $P_\lambda^2$:
\begin{eqnarray}
P_\lambda^2&=&\frac{1}{N^2}\sum_{\sigma,\sigma^\prime\in \lambda}S(\sigma)S(\sigma^\prime)D(\sigma)D(\sigma^\prime)
\nonumber\\
&&=\frac{1}{N^2}\sum_{\sigma,\sigma^\prime\in \lambda}S(\sigma)S(\sigma^\prime)\tau^{\langle \sigma^\prime, \sigma \rangle}\epsilon(\sigma^\prime,\sigma)D([\sigma^\prime+\sigma])
\nonumber\\
&=&\frac{1}{N^2}\sum_{\sigma^{\prime\prime}\in \lambda}\{\sum_{\stackrel{\sigma,\sigma^\prime\in\lambda}{[\sigma^\prime+\sigma]=\sigma^{\prime\prime}}}S(\sigma)S(\sigma^\prime)\tau^{\langle \sigma^\prime, \sigma \rangle}\epsilon(\sigma^\prime,\sigma)\}D(\sigma^{\prime\prime})
\end{eqnarray}
From the subgroup property of $\lambda$: $[\sigma'+\sigma]$ goes
over all of $\lambda$; for given $\sigma,
\sigma^{\prime\prime}\in\lambda$, $\sigma^\prime\in\lambda$ is
unique. The factor $ \tau^{\langle \sigma^\prime, \sigma
\rangle}=\pm 1$. So in the last expression, for each
$\sigma^{\prime\prime}$, $\{\cdots\}$ has exactly $N$ terms, each
$\pm 1$. So, since ${\rm Tr}(P_\lambda)=1$,
\begin{eqnarray}
P_\lambda^2&=&P_\lambda \Longleftrightarrow P_\lambda ~{\rm is~a~rank~one~projection~operator}\nonumber\\
&&\Longleftrightarrow \forall \sigma^{\prime\prime}\in \lambda,
 \frac{1}{N}\sum_{\stackrel{\sigma,\sigma^\prime\in\lambda}{[\sigma^\prime+\sigma]=\sigma^{\prime\prime}}}
S(\sigma)S(\sigma^\prime)\tau^{\langle \sigma^\prime, \sigma \rangle}\epsilon(\sigma^\prime,\sigma)=S(\sigma^{\prime\prime})\nonumber\\
&&\Longleftrightarrow S(\sigma)S(\sigma^\prime)=\tau^{\langle \sigma^\prime, \sigma \rangle}\epsilon(\sigma^\prime,\sigma)S([\sigma+\sigma^\prime]),~\forall \sigma^\prime,\sigma\in\lambda
\label{46}
\end{eqnarray}
If this is obeyed, it implies that
$\{S(\sigma)D(\sigma),~\sigma\in\lambda\}$ give a true
$N$--dimensional UR of $\lambda$. We have to examine : Can these
conditions be imposed consistently for all isotropic lines
$\lambda$? If yes, to what extent are the $S(\sigma)$ then
determined? These are the questions we examine next.
\section {The $N$ odd case}
If $N$ is odd so is $N^2$ and hence from the group structure we have
\begin{eqnarray}
&&\sigma\in\Gamma_0~ {\rm or}~\lambda\Longrightarrow ~\exists ~{\rm unique}~ \sigma^\prime\in\Gamma_0~ {\rm or}~\lambda~{\rm such~ that}~\sigma=[2\sigma^\prime]\nonumber\\
&&i.e., {\rm any}~\sigma=(q,p)=([2q^\prime],[2p^\prime]), ~{\rm unique}~ q^\prime,p^\prime \in \mZ_N.\nonumber\\
&&q~{\rm or}~ p ~{\rm even}, \leq N-1 : q^\prime=q/2\leq(N-1)/2,~p^\prime=p/2\leq (N-1)/2\nonumber\\
&&q~{\rm or}~ p ~{\rm odd} , \leq N-2 : q^\prime=(q+N)/2\geq(N+1)/2,~p^\prime=(p+N)/2\geq( N+1)/2
\end{eqnarray}
In the relation $(\ref{46})$, setting $\sigma=\sigma^\prime$ to get
\begin{equation}
 S([2\sigma^\prime])=\epsilon(\sigma^\prime,\sigma^\prime)
\end{equation}
and looking at $q/p$ even/odd we find that $S(q,p)$ are unambiguously given by :
\begin{equation}
 S(q,p)=(-1)^{qp}
\end{equation}
Does this obey the condition $(\ref{46})$ for all $\lambda$? We find that this is indeed so.
The expression $(-1)^{qp}$ was found by taking $\sigma=\sigma^\prime$ in $(\ref{46})$. Now we put this into that equation with $\sigma$ and $\sigma^\prime$ independent and ask if it is obeyed, i.e. whether or not
\begin{equation}
 \epsilon(\sigma^\prime,\sigma)=S(\sigma^\prime)S(\sigma)S([\sigma^\prime+\sigma])\tau^{\langle\sigma^\prime,
\sigma\rangle}
\end{equation}
holds for all $\sigma^\prime,\sigma\in \lambda$. A key observation here is that since $N$ is odd
\begin{eqnarray}
 &&\langle\sigma^\prime,\sigma\rangle=0 ~{\rm mod}~N \Longrightarrow \langle\sigma^\prime,\sigma\rangle=mN\Longrightarrow \nonumber\\
&&\tau^{\langle\sigma^\prime,\sigma\rangle}=(-1)^m\Longrightarrow \tau^{\langle\sigma^\prime,\sigma\rangle}=(-1)^{mN}=(-1)^{\langle\sigma^\prime,
\sigma\rangle}
\end{eqnarray}
So the question now is whether
\begin{equation}
 \epsilon(\sigma^\prime,\sigma)=(-1)^{q^\prime p^\prime+qp+[q+q^\prime][p+p^\prime]+q^\prime p-p^\prime q}
\label{53}
\end{equation}
We now check the exponent on the RHS in various cases
\begin{equation}
 \begin{array}{cccc}q+q^\prime&p+p^\prime&{\rm exponent}& {\rm rhs}\\
  \leq N-1&\leq N-1&2q^\prime p&+1\\
 \leq N-1&\geq N&N(q^\prime +q)&(-1)^{q^\prime+q}\\
 \geq N&\leq N-1&N(p^\prime +p)&(-1)^{p^\prime+p}\\
 \geq N&\geq N&N^2+N(q^\prime+q+p^\prime +p)&(-1)^{q^\prime+q+p^\prime+p+N}\\
\end{array}
\end{equation}
So, comparing with $\epsilon(\sigma^\prime,\sigma)$, we find that $(\ref{53})$ holds.

Thus in the $N$ odd case the marginals conditions for all isotropic lines can be satisfied, all the $S(\sigma)$ are determined as above.

This unique solution is related to the Fourier matrix (actually parity matrix) results. The Fourier operator $F$ on $\cH$ has these actions and properties
\begin{eqnarray}
&&F|q\rangle=|q),~F|p)=|[-p]\rangle, F^\dagger F=FF^\dagger =F^4=1\\
&&F^2|q\rangle=|[-q]\rangle,~F^2|p)=|[-p]).
\end{eqnarray}
So $F^2=P=$ parity operator, which is what we need. From the relation $(\ref{24})$
\begin{equation}
F|p\rangle\langle q|=\frac{1}{N^{3/2}}\sum_{\sigma^\prime\in\Gamma_0}\omega^{\langle\sigma,\sigma^\prime\rangle+qp}~\tau^{-q^\prime p^\prime}D(\sigma^\prime)
\end{equation}
Set $p=q$ and sum to get
\begin{equation}
F=\frac{1}{N^{3/2}}\sum_{\sigma^\prime\in\Gamma_0}\{\sum_{q\in \mZ_N}\omega^{q^2+q(q^\prime-p^\prime)}\}\tau^{-q^\prime p^\prime}D(\sigma^\prime)
\end{equation}
Next for $P=F^2$, by calculating in the $|q^\prime\rangle$ basis
\begin{eqnarray}
{\rm Tr}(PD(\sigma)^\dagger)=\tau^{-qp}\sum_{q^\prime\in \mZ_N}\omega^{-pq^\prime}\delta_{[q+2q^\prime],0}
\end{eqnarray}
and in the $|p^\prime)$ basis
\begin{eqnarray}
{\rm Tr}(PD(\sigma)^\dagger)=\tau^{qp}\sum_{p^\prime\in \mZ_N}\omega^{qp^\prime}\delta_{[p+2p^\prime],0}
\end{eqnarray}
which are necessarily equal. Using the first form , for $N$ odd,
the Kronecker delta gives
\begin{eqnarray}
q={\rm even} &=&0,2,4,\cdots,N-3,N-1:q^\prime=[N-q/2]\\
q={\rm odd} &=&1,3,5,\cdots,N-2:q^\prime=(N-q)/2\\
{\rm Tr}(PD(\sigma)^\dagger)&=&\begin{cases}q~{\rm even}:\tau^{-qp}\omega^{-p[N-q/2]}=1\\
q~{\rm odd}:\tau^{-qp}\omega^{-p(N-q)/2}=(-1)^p\end{cases}\nonumber\\
&=&(-1)^{qp}
\end{eqnarray}
Hence
\begin{equation}
P=\frac{1}{N}\sum_{\sigma\in\Gamma_0}(-1)^{qp}D(\sigma)
\end{equation}
Now
\begin{equation}
\hW(0,0)=\frac{1}{N}\sum_{\sigma\in\Gamma_0}S(\sigma)D(\sigma)
\end{equation}
and hence
\begin{equation}
W(0,0)=P\Longleftrightarrow S(\sigma)=(-1)^{qp}
\end{equation}
Thus for the case of $N$ odd, we see that there is a unique consistent solution for all signs $S(\sigma)$, such that the marginals conditions can be satisfied for all $\lambda$'s. The resulting Wigner phase point operators are the same as those known in the literature and are characterised by the fact that the phase point operator at the origin is the parity operator as in the continuum case. The existence of a unique square root group element $\sqrt{g(\sigma)}$ for each $g(\sigma)$, guaranteed by $N$ being odd, is adequate for this purpose. In particular it has not been necessary to survey in any sense the set of all isotropic lines $\lambda$, their orbit structure under $SL(2,\mZ_N)$ action (see below) etc.

\section{ The case of $N$ a prime power}

Towards handling the case of general $N$ (essentially even $N$) we
may note the following : Any $N$ can be uniquely written as the
product of powers of (increasing) primes as :
\begin{eqnarray}
 &&N=N_1N_2\cdots N_k=\prod_{j=1}^{k}~N_j\nonumber\\
&& N_j=p_j^{n_j}, ~p_j=j^{\rm {th}}~{\rm prime}:~p_1=2,~ p_2=3,~p_3=5,\cdots,\nonumber\\
&&p_j={\rm odd}~j\geq 2; ~{\rm and}~ n_j=0 ~{\rm or}~ 1 ~{\rm or}~ 2\cdots .
\label{65}
\end{eqnarray}
If $n_1=0$, $N$ is odd and then previous results of Section V  are
in hand. We expect something new to  arise only when  $n_1\geq1$.

We consider the case when $N$ is a power of a single prime in the
rest of this Section, and turn to the general case $(\ref{65})$
later in Section VII. Simplifying the notation as much as possible
for the moment let us write:
\begin{equation}
 N=p^n,~~~p~{\rm prime}, ~n=0,1,2\cdots .
\end{equation}
(Care will be taken to avoid this prime $p$ being confused with
the second entry in the pair $(q,p)$ corresponding to a general
point $\sigma\in \Gamma_0 =\mZ_N\times\mZ_N $).

\subsection{Isotropic Lines  and $SL(2,\mZ_N)$ orbits for $N=p^n$}
For the isotropic lines  we have the following results \cite{14} :
\begin{enumerate}
 \item The total number of isotropic lines $\lambda$ is
\begin{equation}
\cN=(p^{n+1}-1)/(p-1)
\label{67}
\end{equation}

\item The number $\cN(\sigma)$ of isotropic lines passing through
a point $\sigma \in \Gamma_0=\mZ_{N} \times \mZ_{N}$ is computed
as follows. Any $a\in \mZ_N$ can be uniquely written as
\begin{eqnarray}
 a&=& a_0+a_1p+a_2p^2+\cdots+a_{n-1}p^{n-1},\nonumber\\
a_j&\in& \{0,1,\cdots,p-1\},~j=0,1,\cdots,n-1.
\end{eqnarray}
The `$p$--valuation of $a$' is then the smallest $j$ for which $a_j$ is nonzero:
\begin{eqnarray}
v(a)&=& p-{\rm valuation~of}~a\nonumber\\
&=&j~{\rm such~that}~a_0=a_1=\cdots= a_{j-1}=0, a_j\geq 1.
\end{eqnarray}
This definition is unambiguous for $a\geq 1$, in particular, we have:
\begin{eqnarray}
&&v(1)=v(2)=\cdots=v(p-1)=0;\nonumber\\
&&v(p)=1,\cdots~;~v(p^2)=2\cdots~;\cdots~;~v(p^{n-1})=n-1;\nonumber\\
&&v(N-1)=v(p^n-1)=0~{\rm as}~p^n-1=(p-1)(1+p+p^2+\cdots+p^{n-1}).
\end{eqnarray}
We supplement these with the convention
\begin{equation}
 v(0)=n
\end{equation}
based on $p^n=0~{\rm mod}~N$. For $\sigma =(q^\prime,p^\prime) \in
\mZ_n\times\mZ_N$ we define the $p$-valuation by
\begin{equation}
 v(\sigma)=(v(q^\prime),v(p^\prime))_<~,~~~q^\prime,p^\prime\in\mZ_N.
\end{equation}
Thus for instance;
\begin{equation}
 v((0,0))=n;~v((1,p^\prime))=v((q^\prime,1))=v((N-1,p^\prime))=v((q^\prime,N-1))=0.
\end{equation}
Then the number of isotropic lines passing through $\sigma\in\mZ_N\times\mZ_N$ is
\begin{equation}
 \cN(\sigma)=\frac{(p^{v(\sigma)+1}-1)}{(p-1)}
\label{74}
\end{equation}

Comparing with $(\ref{67})$ we see that $\cN=\cN((0,0))$: this is
consistent with the condition that any isotropic line $\lambda$
must contain $(0,0)$. For $\sigma=(1~{\rm or}~N-1,p^\prime),~
(q^\prime,1~{\rm or}~N-1)$ we have $v(\sigma)=0,\cN(\sigma)=1$, so
only one isotropic line passes through each of these points.

\end{enumerate}

We now turn to the group $SL(2,\mZ_N)$
\begin{equation}
 SL(2,\mZ_N) =\{A=\left(\begin{array}{cc}a&b\\c&d\end{array}\right)~|~a, b, c, d\in \mZ_N;~~ad-bc=1~{\rm mod}~N\};
\end{equation}
and its action on $\Gamma_0$ and on the family of isotropic lines
thereof. For the case at hand viz. $N=p^n$ the order
$|SL(2,\mZ_N)|$ is given by
\begin{equation}
 |SL(2,\mZ_N)|=p^{3n-2}(p^2-1)
\end{equation}
It acts on the points and isotropic lines in $\Gamma_0$ as
follows:
\begin{eqnarray}
 A\in SL(2,\mZ_N)&:&\sigma=(q^\prime,p^\prime)\in \Gamma_0\rightarrow \sigma^\prime=(aq^\prime+bp^\prime
 ,cq^\prime+dp^\prime) \in \Gamma_0,\nonumber\\
 &:&\lambda=\{(q^\prime,p^\prime)\}~~~~\rightarrow \lambda^\prime=\{(aq^\prime+bp^\prime,cq^\prime
 +dp^\prime)\}.
\end{eqnarray}
From the latter action one finds that \cite{14}
\begin{enumerate}
\item The $\cN$ isotropic lines,
 divide themselves into $1+[n/2]$ orbits under $SL(2,\mZ_{N})$ action, where $[n/2]$ is the
 integer part of $n/2$ . They are denoted by $O_{k}(p^{n}), k=0,1,\cdots, [n/2]$. For $k< n/2$ the orbit contains
\begin{equation}
\cN(O_k)= (p+1)p^{n-2k-1}
\end{equation}
isotropic lines, while for $k=n/2$ in case $n$ is even we have
\begin{equation}
\cN(O_{n/2})= 1.
\end{equation}
One can easily check in both cases that
\begin{equation}
 \sum_{k,0,1,\cdots}^{[n/2]}\cN(O_k)=\cN
\end{equation}
The largest orbit corresponds to $k=0$ and contains $(p+1)p^{n-1}$ isotropic lines.

\item Only the largest orbit $O_0(p^n)$ has the property that it covers all points
in $\Gamma_0=\mZ_{N}\times \mZ_{N}$. The $(p+1)p^{n-1}$ isotropic lines in this orbit are all
 generated by  single generators of order $N$ which may be taken to be $(1,p^\prime)$ for $p^\prime
 \in\{0,1,\cdots, N-1\}$ and $(q^\prime,1)$ for $q^\prime\in \{0,p,2p,3p,\cdots, (p^{n-1}-1)p\}$.
\end{enumerate}
\subsection{ Isotropic lines in the $2^n$ case}
Now we specialise to the case $N=2^n$, as otherwise $N$ is odd
and then the comprehensive results of Section V are available. Thus in a sense 
this is the most important remaining case. Specialising the above statements 
( and further quoting from \cite{14}) we now have:

\begin{enumerate}
\item  The total number of isotropic lines $\lambda$ is
\begin{equation}
 \cN=2^{n+1}-1=2N-1
\end{equation}
\item  If $\sigma$ is of the form $(2j,2k)$, then from $(\ref{74})$, as $v(\sigma)\geq 1$, the number
of $\lambda$'s passing through it is 
\begin{equation}
\cN(\sigma=(2j,2k))\geq 3.
\end{equation}
If $\sigma$ is of any of the other three forms $(2j,2k+1),(2j+1,2k~{\rm or}~2k+1)$, then as $v(\sigma)=0$ the number 
of $\lambda$'s passing through it is 
\begin{equation}
\cN(\sigma\neq (2j,2k))=1.
\end{equation}
\item  The $\lambda$'s separate into two types 
\\
\noindent
Type (a): $3N/2$ in number, generated by single generators, and comprising a single (the largest) orbit $O_0(2^n)$, 
\\
\noindent
 Type (b) $N/2-1$ in number, involving two generators of orders $2^r$ and $2^s$ with both $r,s$ nonzero and $r+s=n$; and comprising all 
 the remaining  orbits $ O_k(2^n),~k=1,2,\cdots,[n/2]$.
\item  Every phase point $\sigma$ lies on (one or more) $\lambda$'s of Type (a). The $\lambda$'s of type (b) cover all 
the even phase points $(2j,2k)$ only.
\item  The $\lambda$'s of Type (a) separate further into two subtypes:
\\
\noindent
Type (a1) containing $N$ $\lambda$'s generated by $(1,p_0)$ for $p_0 =0,1,2,\cdots, N-1$.
\\
\noindent
Type (a2) containing $N/2$ $\lambda$'s generated by $(q_0,1)$ for $q_0 =0,2,4,\cdots, N-2$

Therefore each $\sigma$ of any of the three types other than $(2j,2k)$ lies on a unique $\lambda$ of Type (a) according to the pattern :
\begin{eqnarray}
&&\sigma=(2j+1, 2k~{\rm or}~2k+1) ---------- {\rm Type~ (a1)}\nonumber\\
&&\sigma=(2j, 2k+1) --------------~{\rm Type~ (a2)}
\end{eqnarray}
\end{enumerate}
\subsection{ Marginals property for Isotropic lines in $\mZ_{2^n}\times \mZ_{2^n}$}
The condition that the average of the phase point operators along an isotropic line $\lambda$ be a one dimensional projector is given in $(\ref{46})$. The other essential conditions on the signs $S(\sigma)$ are the reflection symmetry 
$(\ref{30})$ and the standard marginals conditions $(\ref{31}),(\ref{32})$. We know from $(\ref{37})$ that $\sigma\in \lambda$ implies $[N-\sigma]\in \lambda$ as well. Applying $(\ref{46})$ to such pairs of points on $\lambda$'s of Type (a), and remembering that any $\sigma$ lies on such a $\lambda$, we find that the property $(\ref{30})$ follows.
( For $\lambda$'s of Type (b) this is only partially true as they cover only the even phase points $(2j,2k)$). Thus we begin by imposing only the requirements $(\ref{31}), (\ref{32}), 
(\ref{46})$ on $S(\sigma)$, for all $\lambda$'s of Type (a).

We see from $(\ref{18})$, $N$ being even, that 
\begin{equation}
\epsilon(\sigma,\sigma)=1
\end{equation} 
Therefore setting $\sigma^\prime=\sigma$ in $(\ref{46})$ leads to 
\begin{equation}
S((2j,2k))=1
\end{equation}
This leaves $S((2j+1, 2k~{\rm or}~2k+1))$ and $S((2j, 2k+1))$ to be analysed. Each $\sigma$ of the former type is on a unique Type (a1) $\lambda$, while each $\sigma$ of the latter type is on a unique Type (a2) $\lambda$. We 
apply $(\ref{46})$ in these cases, choosing $\sigma=\sigma_0=(1,p_0)~{\rm or}~(q_0,1)$ and $ \sigma^\prime=[2j\sigma_0]$
or $[2k\sigma_0]$ respectively, thus reaching all points $\sigma$ other than $(2j,2k)$, and relating $S$ at such points to $S(\sigma_0)$:
\begin{eqnarray}
S((2j+1,[(2j+1)p_0]))&=&(-1)^{(2jp_0-[2jp_0])/N}\cdot S((1,p_0))\times\nonumber\\
&&\begin{cases}~~1~{\rm if}~p_0+[2jp_0]\leq N-1,\\-1~{\rm if}~p_0+[2jp_0]\geq N\end{cases};
\end{eqnarray}
\begin{eqnarray}
S(([(2k+1)q_0],2k+1))&=&(-1)^{(2kq_0-[2kq_0])/N}\cdot S((q_0,1))\times\nonumber\\
&&\begin{cases}~~1~{\rm if}~q_0+[2kq_0]\leq N-1,\\-1~{\rm if}~q_0+[2kq_0]\geq N.\end{cases}
\end{eqnarray}
In the former relation, the choice of $2k$ or $2k+1$ determines $p_0$ uniquely ; in the latter, that of $2j$ determines $q_0$ uniquely. For $q_0=0$, eq. $(\ref{31})$ determines $S((0,1))=1$; for $p_0=0$, eq. $(\ref{31})$ determines $S((1,0))=1$. The remaining $3N/2-2$ undetermined signs are $S((q_0,1)), ~q_0=2,4,\cdots, N-2$ and 
$S((1,p_0)), ~p_0=1,2,\cdots, N-1$.

These conditions may equivalently be written  as 
\begin{eqnarray}
S((2j+1,[(2j+1)p_0]))&=&S((1,p_0))\times
\begin{cases}~~~~~~1~{\rm if}~((2j+1)p_0-[(2j+1)p_0])/N {\rm~is~even}\\
~~-1~{\rm if}~((2j+1)p_0-[(2j+1)p_0])/N {\rm~is~ odd}
\end{cases}\\
S(([(2k+1)q_0,(2k+1))&=&S((q_0,1))\times
\begin{cases}~~~~~~1~{\rm if}~((2k+1)q_0-[(2k+1)p_0])/N {\rm~is~even}\\
~~-1~{\rm if}~((2k+1)q_0-[(2k+1)q_0])/N {\rm~is~odd}\end{cases}.
\end{eqnarray}
Thus, for $n=1,2$, when $N=2$ and $4$ respectively, the free signs are indicated thus:
\begin{equation}
\begin{array}{ccccccc}S(q,p)  &&\\
                                                           & &1&1  \\
                                                          & &1&S(1,1)\end{array}\\
                                                 \end{equation}

\begin{equation}
\begin{array}{ccccccc}S(q,p) & & && \\
                             && 1&1 &1&1 \\
                            && 1& S(1,1)& S(1,2)&S(1,3) \\
                             &&1 & S(2,1)& 1& -S(2,1)\\
                              &&1& S(1,3)&-S(1,2) & S(1,1)\end{array}
\end{equation}
In summary: the marginals conditions can be consistently imposed on all isotropic lines of Type (a) comprising the largest orbit but leaving $3\times 2^{n-1}-2$ of the $S(\sigma)$ unfixed.

We next see by low dimensional examples that these conditions cannot be consistently extended to include isotropic lines of type (b). For $n=1, N=2$, there are no isotropic lines of type (b).
For $n=2,N=4$, there is one isotropic line of type (b),  generated by $(2,0)$~and~$(0,2)$.
Condition $(\ref{46})$  when applied to this isotropic line gives $ S(2,2)=-1$ conflicting with $S(2,2)=1$ obtained from isotropic lines of type (a). For $n=3, N=8$, there are three isotropic lines  of type (b) generated respectively by $\{(2,0),(0,4)\}$,$\{(0,2),(0,4)\}$,$\{(2,2),(0,4)\}$ . Again, as for $n=2,N=4$ one finds that the results of $(\ref{46})$ for isotropic lines of type (b) conflict with those for isotropic lines of type (a)-- one can not impose  marginals property on all isotropic lines consistently.

\section{Isotropic Lines and orbits in the general case}
Turning now to the case of a general $N$ we note that ring $\mZ_N$ can be factored as
\begin{equation}
\mZ_N=\mZ_{N_1}\times \mZ_{N_2} \times \cdots \times \mZ_{N_k}
\end{equation}
The explicit correspondence between elements of $\mZ_N$ and those of the rings $\mZ_{N_j}$ is provided by the chinese remainder theorem which tells us that an element $q\in \mZ_N$ can be uniquely decomposed as
\begin{equation}
q= \sum_{j=1}^{k}q_j\cdot \nu_j \cdot \mu_j
\end{equation}
where $q_j=[q ~{\rm mod}~N_j] \in \mZ_{N_j}$, $\nu_j =N/N_j$ and $\mu_j$ denotes the (multiplicative) inverse of $\nu_j$ in $\mZ_{N_j}$. Thus each element $q\in \mZ_N$ can be uniquely represented as an array
\begin{equation}
q\longleftrightarrow \{q_1,q_2,\cdots,q_k\}, ~~q_i\in \mZ_{N_i}
\end{equation}
In particular the elements $0$ and $1$ are represented by
\begin{eqnarray}
0\longleftrightarrow \{0,0,\cdots,0\};~~
1\longleftrightarrow \{1,1,\cdots,1\}
\label{78}
\end{eqnarray}
Further, this correspondence has the nice property that
\begin{eqnarray}
q+q^\prime &\longleftrightarrow& \{q_1+q_1^\prime,q_2+q_2^\prime,\cdots,q_k+q_k^\prime\}, ~~q_i~{\rm and}~q_i^\prime\in \mZ_{N_i}\nonumber\\
\\
qq^\prime &\longleftrightarrow& \{q_1q_1^\prime,q_2q_2^\prime,\cdots,q_kq_k^\prime\}, ~~q_i~{\rm and}~q_i^\prime\in \mZ_{N_i}\nonumber
\label{68}
\end{eqnarray}
In view  of this and the properties $(\ref{78})$ and $(\ref{68})$ we have the following results:
\begin{itemize}
 \item
A point $\sigma\in \mZ_N\times \mZ_N$ can be represented as
\begin{equation}
\sigma \longleftrightarrow \{\sigma_1,\sigma_2,\cdots,\sigma_k\},~~\sigma_i\in \mZ_{N_i}\times \mZ_{N_i}
\end{equation}
\item The symplectic product  $\langle \sigma,\sigma^\prime\rangle$ vanishes if and only if each of the components $\langle \sigma_i,\sigma_i^\prime\rangle$ vanish.
\item The group $SL(2,\mZ_N)$ also factorises as
 \begin{equation}
 SL(2,\mZ_N)=SL(2,\mZ_{N_1})\times SL(2,\mZ_{N_2})\times SL(2,\mZ_{N_k})
\end{equation}
This can easily be seen by considering the case $N=N_1N_2$ and verifying that any matrix $A\in SL(2,\mZ_N)$ :
\begin{equation}
 A=\left(\begin{array}{cc}a&b\\c&d\end{array}\right);~ab-cd=1;~~a, b, c, d\in \mZ_n;
\end{equation}
can be decomposed as $A=A_1A_2$ where
\begin{equation}
 A_1=\left(\begin{array}{cc}(a_1,1)&(b_1,0)\\(c_1,0)&(d_1,1)\end{array}\right)\in SL(2,\mZ_{N_1});~
A_2=\left(\begin{array}{cc}(1,a_2)&(0,b_2)\\(0,c_2)&(1,d_2)\end{array}\right)\in SL(2,\mZ_{N_2});
\end{equation}
\end{itemize}
From these considerations it is evident  that the isotropic lines in
 $\sigma\in \mZ_N\times \mZ_N$ and $SL(2,\mZ_N)$ action are completely determined by those in each of the factors $\mZ_{N_j}\times \mZ_{N_j}$.

\section {Tomographic reconstruction for $N=2^n$}

From the discussion towards the end of Section VI it is evident that for the case when $N=2^n$  we can only insist on  marginals property restricted to the isotropic lines of Type (a) constituting the largest orbit  under $SL(2,\mZ_N)$ action. For each choice for  the free signs, we can associate with each such isotropic line a rank one projector $P_\lambda$. Each isotropic line generates $N-1$ other lines 
`parallel' to it obtained, for instance, by shifting the points on it by an amount $(0,i)$ in the case of isotropic lines of Type a1 and by an amount $(i,0)$ in the case of isotropic lines of Type a2 with $i$ taking values $1, 2,\cdots, N-1$. Denoting by $(\lambda,i)$ the lines parallel to the line $\lambda$, the projectors  $P_{\lambda,i}$  associated with them are obtained by the conjugate action of the appropriate  displacement operators on $P_{\lambda}$: 
\begin{equation}
P_{(\lambda, i)}=\begin{cases}&D(0,i)P_{\lambda} D^\dagger(0,i)~{\rm if~\lambda ~is~ of~ Type~a1}\\
                      & D(i,0)P_{\lambda} D^\dagger(i,0)~{\rm if~\lambda ~is~of ~Type~a2 }\end{cases}\\
                       \end{equation}
From  these $N$ projectors associated with each isotropic line we can construct $N-1$ traceless operators $T_{(\lambda,i)}=P_{(\lambda,i)}-I_N/N$. Each line of Type (a) gives us $N-1$ $T$'s and since there are $3N/2$ lines of Type (a), we have a collection of $3N(N-1)/2$ traceless hermitian operators. Given a density operator for an $N$-state system, the operator 
$\rho_N-I_N/N$ belongs to the $N^2-1$ dimensional real Hilbert space of $N\times N$ traceless Hermitian matrices. The question concerning the tomographic reconstruction of $\rho$ then reduces to the question  whether or not the collection of the $T$'s  above spans the $N^2-1$ dimensional real vector space of traceless hermitian operators. This can be checked by examining the rank of the Gram matrix associated with the $T's$. For $N=2,4$ we have explicitly checked that the corresponding Gram matrices indeed have ranks 3 and 15 respectively. Thus, it would seem that even with restricted marginals property the construction developed here permits a tomographic reconstruction of the state of an $N$-level system though in a non optimal fashion -- we have $(N-1)(N-2)/2$ more $T$'s then the $N^2-1$ required. 

\section{Concluding Remarks}
 We have shown how to set up Wigner distributions for finite even dimensional quantum systems working entirely with a $N\times N$ lattice instead of
  a $2N\times 2N $ grid as was found necessary in the existing formalisms. The Wigner distributions thus obtained are consistent with a restricted
  marginals property and are characterised by $3.2^{n-1}-2$. undetermined signs  where $n$ is the exponent of $2$ in the decomposition of $N$ into
  prime factors. As a result, for instance, for $N=2,4,8,16$ there are $2,2^4,2^{10}, 2^{22}$ different possible definitions of Wigner distributions. 

As a curiosity, in the spirit of the work in \cite{15}, we have also examined the dependence of the eigenvalues of the phase point operators for  $N=2, 4, 8$, as a function of the signs that remain free. (For this purpose it is sufficient to look at the eigenvalues of ${\widehat W}(0,0)$). We find that:

For $N=2$ there is only one free sign, $S(1,1)$, and the spectrum of ${\widehat W}(0,0)$ is the same for
$S(1,1)=\pm 1$

For $N=4$  there are three distinct spectra for ${\widehat W}(0,0)$  depending on the values of the four free signs, $S(1,1)\equiv a, S(1,2)\equiv b,S(1,3)\equiv c, S(2,1)\equiv d$. They are
\begin{itemize}
\item $((1+\sqrt{6})/2,(1-\sqrt{6})/2,-1/2,1/2)$ corresponding to $a=c,b=-d$ and $a=-c,b=d$.
\item  $((1+2\sqrt{2})/2,-1/2,(1-\sqrt{2})/2,(1-\sqrt{2})/2$ corresponding to $a=c=1,b=d=1$,$a=c=-1,b=d=-1$ and
$a=-c=1,b=-d$
\item  $((1+\sqrt{2})/2,(1+\sqrt{2})/2,(1-2\sqrt{2})/2,-1/2)$ corresponding to $a=c=1,b=d=-1$,$a=c=-1,b=d=1$ and
$a=-c=-1,b=-d$
\end{itemize}

For $ N=8$, one has $4$ and $N=16$, one has $15$ distinct spectra. Thus, although the number of different Wigner distributions based on choices for the signs for $N=2,2^2,2^3,2^4$ is $2,2^4,2^{10}, 2^{22}$, those which have distinct spectra are only $1,3,4,15$ in number.
( It seems that the number of distinct spectra for $N=2^n$ equals $2^{n-1}$ if n even and $2^n-1$ if n odd)
The question as to what bring about this
  enormous reduction is under investigation. Further, it would  be interesting to see if the square root idea developed here works in the case when the coordinates take values in a finite field \cite{5} and to see how it relates to Wigner distributions in the more general setting based on the theory of frames employed in the work of Ferris and Emerson \cite{F}.
\vskip5mm
\noindent
{\bf Acknowledgements} We are extremely grateful to Olivier Albouy, Maurice Kibler and Michel Planat for their readiness to help whenever their help was sought. One of (SC) would also like to acknowledge many fruitful converations with David Gross, Macus Appleby and Ingemar Bengtsson. 

\end{document}